\documentstyle[aps,prd]{revtex}

\newcommand{\be}{\begin{equation}}
\newcommand{\ee}{\end{equation}}
\newcommand{\ba}{\begin{eqnarray}}
\newcommand{\ea}{\end{eqnarray}}

\begin{document}
\title{Enlarged geometries of gauge bundles}
\author{R. Aldrovandi and A. L. Barbosa}
\address{Laboratoire de Gravitation et Cosmologie Relativistes \\
Universit{\'e} Pierre et Marie Curie, CNRS/ESA 7065 \\
4, Place Jussieu \ Cedex 05 \\
75252 Paris, France}
\date{\today}
\maketitle

\begin{abstract}

The geometrical picture of gauge theories must be enlarged when a gauge
potential ceases to behave like a connection, as it does in electroweak
interactions. When the gauge group has dimension four, the vector space
isomorphism between spacetime and the gauge algebra is realized by a
tetrad--like field. The object measuring the deviation from a strict bundle
structure has the formal behavior of a spacetime connection, of which the
deformed gauge field--strength is the torsion. A generalized derivative
emerges in terms of which the two Bianchi identities are formally recovered. 
Effects of gravitational type turn up.
The dynamical equations obtained correspond to a broken gauge model on a
curved spacetime.

\end{abstract}

\section{Introduction}
\label{Introduction}

Differential Geometry, in its modern fiber-bundle language, provides 
the mathematical background for the theories describing the known 
fundamental interactions.  The bundle of frames stands behind General 
Relativity, while other principal bundles, built up with the 
respective gauge groups, give a clear picture of the kinematic setup 
backing electroweak and strong interactions (see Trautman, 1970; Wu 
and Yang, 1975; Daniel and Viallet, 1980).  The picture is nowadays 
commonplace: geometry supplies the stage-set, on which Lagrangians of 
phenomenological origin rule over dynamics.  Dynamics confers 
different characters to gravitation, whose Lagrangian is of first 
order in the curvature, and to the other interactions, whose 
Lagrangians are of second order in the curvature.  But in all cases it 
is a curvature which appears, and curvature is a quantity derived from 
a connection.  The metric keeps the central role in gravitation, but 
the basic fields in the other cases are gauge potentials, that is, 
connections.  A splendid experimental record favors the existing 
theories and justifies the belief that much of their content is of 
perennial value.

There are, however, some cloudy spots in this sunny landscape. There are too
many arbitrary constants and a obstinate lack of unity with respect to great
general principles. Gravitation alone is universal, can be locally simulated
by a moving frame, has a problematic energy and is power--counting
non--renormalizable. Some of the mediating bosons are massless and have
problematic charges. Other are massive and have well-defined charges. And
there is the question of the meaning to be attributed to spontaneous
symmetry breakdown. The presence of a remnant scalar field adds to the
difficulties (Gaillard, Grannis and Sciulli, 1999). This blending of experimental success and
theoretical bafflement suggests that, though the gauge principle is promised
an important role in an eventual final theory, the simple, direct gauge
prescription will not have the last saying as it stands. In the search of a
more comprehensive framework, string theory, with its ultimate goal of
explaining, in principle, ``everything'', is the dominating trend.

We want to present here a few more steps of another proposal 
(Aldrovandi, 1995), which starts from gauge theories and looks for the minimal 
modifications necessary to enlighten at least some of these 
difficulties.  It takes into account two initial clues.  The first is 
supported by all the experimental evidence and is concerned with the 
peculiar behavior of the electroweak gauge potentials.  The gauge 
potentials appearing in chromodynamics and isolated electromagnetism, 
as well as the Christoffel symbols in gravitation, behave strictly as 
connections, but the vector fields describing real particles in 
electroweak theory do not.  The theory does start with a 
connection-behaving gauge potential, but then spontaneously breaks the 
symmetry by introducing an external field.  The final combinations, 
representing the physical fields, do not transform as connections.  
This leads to the second clue, more mathematical in nature: when a 
gauge field ceases to behave like a connection, the whole geometric 
picture provided by the underlying bundle is blurred.  What happens to 
the bundle picture when a connection, or part of it, adopts an 
abnormal behavior ?

The connection adjoint behavior is essential to the bundle picture. On the
bundle tangent spaces, it is reflected in the commutators of the vector
fields coming from the base manifold (external space, spacetime) and from
the structure (internal, gauge space) group. Vector fields are derivatives,
and a connection allows mixing internal and external vectors to produce more
general, covariant derivatives, while preserving the bundle makeup. This
preservation comes from the connection adjoint behavior. Any deviation from that
behavior changes the whole picture, and the electroweak physical fields do
deviate. Some encouraging results have been obtained years ago, in which an
abnormal behavior of a gauge potential was shown to engender fields
strongly suggestive of linear connections with their curvatures and
torsions (Aldrovandi 1, 1991), hinting thereby to a relationship with gravitation.
We intend here to present some new results on the subject, valid when the
gauge group has, as in the Weinberg--Salam theory,  dimension 4.

We start (section \ref{ext}) with a formal compact on Lie algebra 
extensions.  In this section we also examine the behavior of the Lie 
algebra of fields on a manifold under changes of basis.  To alleviate 
notation we shall, as a rule, omit projections, their differentials 
and corresponding pull--backs.  In section \ref{Changesofbasis} we 
introduce an enlarged concept of change of basis in the principal 
fiber bundle and we apply them successively to the simplest 
conceivable geometric configuration and obtain 3 kinds of commutation 
relations: those of a gauge theory, those of an extended gauge theory 
and those of a gravitational model.  Non-covariant derivatives, akin 
to those appearing in electroweak theory, turn up naturally in the 
extended formalism.  In Section \ref{recovering} we begin discussing 
which aspects of the geometric picture can still be retained in the 
presence of anomalous connections.  When the base manifold and the 
gauge group have the same dimension, as is the case involving 
spacetime and the electroweak theory, a tetrad--like field is 
naturally introduced to represent the isomorphism of the underlying 
vector spaces of the tangent field algebras.  The object measuring the 
breaking of the bundle structure acquires the aspect of an external, 
linear connection, which preserves the metric defined by the tetrad 
and is endowed with curvature and torsion.  Thus, the same {\em 
objects} of usual geometry are found and strongly suggest a relation 
to gravitation.  Section \ref{EqsandIdents} is devoted to show that 
such objects have the expected geometrical properties and lead to 
reasonable dynamical equations.  Many results previously found for the 
translation group (Aldrovandi 2, 1991) are extended to the non-abelian 
case.  It should be emphasized that, due to the presence of 
non--covariant objects, even the most trivial formulas of tensor 
calculus must be reworked from the start.  Some of them survive, other 
appear modified.  A very general derivative shows up, involving 
simultaneously gauge and ``gravitational'' aspects.  Dynamics for the 
gauge sector can be obtained by assuming the persistence of the 
duality symmetry and, for the gravity sector, by a procedure analogous 
to that used in General Relativity.  The final section sums up the 
results and the many still unsolved problems.

%%%%%%%%%%%%%%%%%%%%%%%%%%%%%%%%%%%%%%%%

\section{Extensions of tangent algebras}
\label{ext} 
%
%                                                     %
%%%%%%%%%%%%%%%%%%%%%%%%%%%%%%%%%%%%%
We shall find it necessary to call attention to a certain number of
elementary facts, and profit to introduce notation through an overview
of well-known notions.  A Lie algebra is a vector space on which a
binary internal operation is defined which is antisymmetric and
satisfies the Jacobi identity.  The operation will be indicated by the
commutator and the algebra whose underlying vector space is $V$ will
be denoted $V^{\prime }$.  For simplicity, the same notation will be
used for a Lie group $G$ and its Lie algebra $G^{\prime }$.  The
algebra is characterized by the operation table written in a vector
basis $\{Y_{\alpha }\}$ of members, $\left[ Y_{\alpha },Y_{\beta
}\right] _{V}$ = $%
f^{\gamma }{}_{\alpha \beta }Y_{\gamma }$. The numbers $f^{\gamma
}{}_{\alpha \beta }$ are the structure constants of $V^{\prime }$. We shall
sometimes indicate the algebra by one of its basis, as in $V^{\prime }$ = $%
\{Y_{\alpha }\}$.

In order to discuss the extension of a Lie algebra $L^{\prime}$ by another
Lie algebra $V^{\prime}$, notice beforehand that the direct sum $E = L
\oplus V$ of the underlying vector spaces $L$ and $V$ is always defined. To
extend $L^{\prime}$ by $V^{\prime}$ means, in general terms, to give an
answer to the following question: when and how can we combine $L^{\prime}$
and $V^{\prime}$ to build another Lie algebra $E^{\prime}$ with underlying
vector space $L \oplus V$ ? In the generic case, many answers are possible,
provided $L^{\prime}$ has a representation acting on $V^{\prime}$. Two main
points should be specified: (i) the insertion of the algebras in the
enlarged space $E$ and (ii) the relationship between the algebras after the
insertion.

We shall be interested in extensions involving the algebras of vector fields
on differentiable manifolds. The pattern introduced below is closely related
to that present on the total manifold $P$ of a principal bundle 
(Kobayashi and Nomizu, 1963).

Let $P$ be a differentiable manifold. It will have a tangent space $T_{p}P$
at each point $p \in P$. A vector field $X$ is a differentiable choice of a
vector $X_{p}$ at each $T_{p}P$. In general, making such a choice is only
possible locally, that is, on an open neighborhood of each point $p$. For
that reason all the discussion which follows will be purely local in
character. If the manifold is $C^{\infty}$, $X$ will act on a space ${\bf \ R%
}(P)$ of infinitely differentiable real functions on $P$.

The set of all vector fields on $P$ constitutes an infinite Lie algebra $\Xi
(P)$. Consider a Lie group whose Lie algebra $G^{\prime}$ has generators $%
J_{\mu}$ satisfying the commutation rules

\begin{equation}
\left[J_{\mu}, J_{\nu}\right] = f^{\lambda}{}_{\mu \nu} J_{\lambda} \; .
\end{equation}

When $G$ acts on $P$ as a transformation group, there is a representation $%
\rho$ of its generators by fields on $P$.  This means (Aldrovandi and 
Pereira, 1995) that
$\rho$ chooses, for each $J_{\mu}$, a {\em representative} field
$Y_{\mu} \in \Xi (P)$:
\[
\rho: G^{\prime} \rightarrow \Xi (P) 
\]
\begin{equation}
\rho: J_{\mu} \rightarrow Y_{\mu} = \rho (J_{\mu}) \; .
\end{equation}
The representation $\rho$ will be a {\it linear representation} when the
representative fields have the same commutation rules as the fields they
represent:
\begin{equation}
\left[Y_{\mu}, Y_{\nu}\right]_{\Xi (P)} = f^{\lambda}{}_{\mu \nu}
Y_{\lambda} \; . \label{YY}
\end{equation}
Suppose that a first representative algebra $L^{\prime}$ = $\{ Y_{\mu} \}$
is given around a point $p$ on $P$, with a number $d \; < \; $ dim $P$ of
generators. Consider also a linear representation, also around $p$, of
another algebra $V^{\prime}$, locally given by a number $n$ = dim $P - d$ of
fields $X_{a}$ with commutations
\begin{equation}
\left[X_{a}, X_{b}\right]_{\Xi (P)} = f^{c}{}_{a b} X_{c} \; .
\end{equation}
If all the involved fields $Y_{\mu}$ and $X_{a}$ are linearly independent,
the set $\{ Y_{\mu}, X_{a} \}$ constitutes a local basis around $p$. Notice
that, once an algebra is represented by vectors at a point $p \in P$, its
structure {\em constants} can become point-dependent (structure {\em
coefficients}) when these vectors are extended into vector fields
around $p$.  As a last basic assumption, suppose the commutation table
in that basis to have the form
\[
\left[Y_{\mu}, Y_{\nu}\right]_{\Xi (P)} = f^{\lambda}{}_{\mu \nu}
Y_{\lambda} - \beta^{a}{}_{\mu \nu} X_{a} \; ; 
\]
\begin{equation}
\left[Y_{\mu}, X_{a}\right]_{\Xi (P)} = C^{b}{}_{\mu a} X_{b} \; ;
\label{table1}
\end{equation}
\[
\left[X_{a}, X_{b}\right]_{\Xi (P)} = f^{c}{}_{a b} X_{c} \; . 
\]
$\beta_{\mu \nu}$ is a 2-form with values in the $V^{\prime}$ sector. 
It characterizes the deviation from the linearity (\ref{YY}) of the
algebra $\{ Y_{\mu} \}$, caused by its association with the algebra
$\{ X_{a} \}$.  The latter, by the above relations, is unaffected: it
is simply included in $E^{\prime}$, and its structure coefficients
remain constant:
\[
\left[X_{a}, X_{b}\right]_{\Xi (P)} = \left[X_{a}, X_{b}\right]_{V} =
f^{c}{}_{a b} X_{c} \; . 
\]
The middle expression in (\ref{table1}) says that the result of any
action of $L^{\prime}$ on $V^{\prime}$ stays in $V^{\prime}$.  For
each fixed $\mu$, the field $Y_{\mu}$ is represented on the $X_{a}$'s
by the matrix $C_{\mu}$ whose entries are the coefficients
$C^{b}{}_{\mu a}$.  The algebra $E^{\prime}$ specified by
(\ref{table1}) is an extension of the representative field algebra of
$L^{\prime}$ by the representative field algebra of $V^{\prime}$.

An extension is {\it trivial} when there is no departure from linearity,
that is, when $\beta^{a}{}_{\mu \nu} = 0$. The extension is a {\it \
direct--product} when the fields $Y_{\mu}$ act on the $X_{a}$'s by the null
representation, that is, when $C^{b}{}_{\mu a} = 0$. This will be a
necessary (but not sufficient) condition for the geometry of gauge theories.

The compound so obtained depends, thus, on the pair $(C^{b}{}_{\mu
a},\beta^{a}{}_{\mu \nu})$.  The extended algebra should be a Lie
algebra, so that we impose the Jacobi identities on the fields obeying
(\ref{table1}).  Three conditions come out, which must be respected by
any pair $(C^{b}{}_{\mu a},\beta^{a}{}_{\mu \nu})$:
\[
Y_{\mu}(\beta^{a}{}_{\nu \sigma}) + Y_{\sigma}(\beta^{a}{}_{\mu \nu}) +
Y_{\nu}(\beta^{a}{}_{\sigma \mu }) + C^{a}{}_{\nu c} \beta^{c}{}_{\sigma
\mu} + C^{a}{}_{\sigma c} \beta^{c}{}_{\mu \nu} + C^{a}{}_{\mu c}
\beta^{c}{}_{\nu \sigma} \ \ \ \ \ \ 
\]
\begin{equation}
\; \; \; \; \; \; \; \; \; \; \; \; \; \; \; \; \; \; \; \; \; \; \; 
\; \; \; \; \; \; \; \; \; \; \; \; \; \; \; \; \; + \ f^{\rho}{}_{\mu 
\nu} \beta^{a}{}_{\sigma \rho} + f^{\rho}{}_{\sigma \mu} 
\beta^{a}{}_{\nu \rho} + f^{\rho}{}_{\nu \sigma} \beta^{a}{}_{\mu 
\rho} = 0 \; ;
\label{Jacobi1}
\end{equation}
\[
Y_{\mu}(C^{a}{}_{\nu b}) - Y_{\nu}(C^{a}{}_{\mu b}) + C^{a}{}_{\mu c}
C^{c}{}_{\nu b} - C^{a}{}_{\nu c} C^{c}{}_{\mu b} - f^{\rho}{}_{\mu \nu}
C^{a}{}_{\rho b} \; \; \; \; \; 
\]
\begin{equation}
\; \; \; \; \; \; \; \; \; \; \; \; \; \; \; \; \; \; \; \; \; \; \; 
\; \; \; \; \; \; \; \; \; \; \; \; \; \; \; \; \; \; \; \; \; \; \; 
\; \; \; \; \; \; \; \; \; \; \; \; - \ 
X_{b}(\beta^{a}{}_{\mu \nu}) - \beta^{c}{}_{\mu \nu} f^{a}{}_{b c} = 0 
\; ;
\label{Jacobi2}
\end{equation}
\begin{equation}
X_{a}(C^{c}{}_{\mu b}) - X_{b}(C^{c}{}_{\mu a}) - C^{d}{}_{\mu a} f^{c}{}_{b
d} + C^{c}{}_{\mu d} f^{d}{}_{b a} + C^{d}{}_{\mu b} f^{c}{}_{a d} = 0 \; .
\label{Jacobi3}
\end{equation}

An extension is {\it central} when $\beta^{c}{}_{\mu \nu} X_{c}$ has all its
elements in the center of the algebra $V^{\prime}$. In particular, it
follows from (\ref{Jacobi2}) that every direct product ($C^{a}{}_{\mu b}$ =
0) is a central extension. In effect, in that case
\begin{equation}
\left[X_{b}, \beta^{a}{}_{\mu \nu} X_{a}\right]_{\Xi (P)} = \{
X_{b}(\beta^{c}{}_{\mu \nu}) + f^{c}{}_{b a} \; \beta^{a}{}_{\mu \nu} \} \;
X_{c} = 0 \; .  \label{central}
\end{equation}

Let us examine what happens to the above commutation tables under a
change of basis.  Starting from the basis $\{ Y_{\mu}, X_{a} \}$ on
the whole manifold $P$, we introduce the particular transformationss
\begin{equation}
Y_{\mu }^{\prime }=Y_{\mu }\ -\ \;\alpha^{c}{}_{\mu }(x)X_{c}\;,
\label{change1}
\end{equation}
where the $\alpha^{a}{}_{\mu }$'s are point-dependent objects on $P$. In the
applications we have in mind $P$ will be the total space of a bundle with
spacetime as base manifold. The fields $Y_{\mu }$ will represent
translations on spacetime, so that the coefficients $f^{\rho }{}_{\mu \nu }$
are a mere signal of anholonomy. We shall take for $\{Y_{\mu }\}$ a
holonomic basis, so that $f^{\rho }{}_{\mu \nu }=0$ in (\ref{table1}), (\ref
{Jacobi1}) and (\ref{Jacobi2}). Notice that the non-linearity indicator $%
\beta ^{a}{}_{\mu \nu }$ is not an anholonomy coefficient for $\{Y_{\mu }\}$%
, as it points toward other directions in the algebra.  Notice that we 
consider (\ref {change1}) as a change of basis on the whole local 
algebra of vector fields on $P$.  The new set of commutation relations 
is (we also drop the index `` $%
\Xi (P)$'' from now on): 
\begin{equation}
\left[ Y^{\prime }{}_{\mu },Y^{\prime }{}_{\nu }\right] =-\;\beta ^{\prime
a}{}_{\mu \nu }X_{a}\;;  \label{newcommut1}
\end{equation}
\begin{equation}
\left[ Y^{\prime }{}_{\mu },X_{a}\right] =C^{\prime b}{}_{\mu a}X_{b}\;;
\end{equation}
\begin{equation}
\left[ X_{a},X_{b}\right] =f^{c}{}_{ab}X_{c}\;,
\end{equation}
with new coefficients given by 
\begin{equation}
C^{\prime b}{}_{\mu a}=C^{b}{}_{\mu a}\;-\;\alpha ^{c}{}_{\mu
}f^{b}{}_{ca}+X_{a}(\alpha ^{b}{}_{\mu }),  \label{CC'}
\end{equation}
\begin{equation}
\beta ^{\prime a}{}_{\mu \nu }=\beta ^{a}{}_{\mu \nu }+K^{a}{}_{\mu
\nu }\;,
\label{betaprime1}
\end{equation}
and
\[
K^{a}{}_{\mu \nu }=Y^{\prime }{}_{\mu }\alpha ^{a}{}_{\nu }-Y^{\prime
}{}_{\nu }\alpha ^{a}{}_{\mu }+\alpha ^{b}{}_{\mu }X_{b}(\alpha
^{a}{}_{\nu })-\alpha ^{b}{}_{\nu }X_{b}(\alpha ^{a}{}_{\mu })
\]
\begin{equation}
\; \; \; \; \; \; \; \; \; \; \; \; \; \; \; \; \; \; \; \; \; \; \; 
\; \; \; \; \; \; \; \; \; \; \; \; \; \; \; \; \; \; \; \; \; \; \; 
 +\ \alpha 
^{b}{}_{\nu }C^{\prime a}{}_{\mu b}-\alpha ^{b}{}_{\mu }C^{\prime 
a}{}_{\nu b}+f^{a}{}_{bc}\alpha ^{b}{}_{\mu }\alpha ^{c}{}_{\nu }\;.
\label{K}
\end{equation}
Relations (\ref{CC'})-(\ref{K}) are such that the forms of the Jacobi
identities are preserved under (\ref{change1}). This is important, 
because as we shall see later, the field equations will come from 
Jacobi Identities.

%%%%%%%%%%%%%%%%%%%%%%%%%%%%%%%%%%%%%%%%

\section{Changes of basis}
\label{Changesofbasis}
%%%%%%%%%%%%%%%%%%%%%%%%%%%%%%%%%%%%%%%%

The simple scheme of basis transformation presented above can, if we start 
from a trivial initial algebra, engender 3 types of algebra: that
of a gauge theory, the extension given above or the forthcoming
extended algebra of section \ref{recovering}, and the algebra
corresponding to a gravitational model.  Assuming the validity of the
duality prescription applied to the Bianchi identities, we can also
obtain the corresponding dynamics of each theory.

We shall take as starting field configuration that appearing on a
fiber bundle whose structure group $G$ has Lie algebra $G^{\prime }$ =
$\{X_{a}\}$, and whose base manifold is spacetime represented by the
trivial holonomic basis $\{\partial _{\mu }\}$.  The set of
commutation relations is
\[
\left[ \partial _{\mu },\partial _{\nu }\right] =0\;;\;\;\;\;\;\;\;\;\; 
\]
\begin{equation}
\left[ X_{a},\partial _{\mu }\right] =0\;;\;\;\;  \label{config1}
\end{equation}
\[
\left[ X_{a},X_{b}\right] =f^{c}{}_{ab}X_{c}\;. 
\]
It represents a trivial and direct--product extension of the
translation algebra by $G^{\prime }$, or vice-versa.  Physically, it
corresponds to a theory without interaction.

Let us first make in (\ref{config1}) a change of basis 
\begin{equation}
X_{\mu }=\partial _{\mu }\ -\ \;\alpha ^{a}{}_{\mu }X_{a}\;,  \label{change2}
\end{equation}
imposing that it preserves the direct--product character. It leads to
\[
\left[ X_{\mu },X_{\nu }\right] =-\;\beta ^{a}{}_{\mu \nu }X_{a}; 
\]
\begin{equation}
\left[ X_{a},X_{\mu }\right] =0\;;\;\;\;\;\;\;  \label{dirprod}
\end{equation}
\[
\left[ X_{a},X_{b}\right] =f^{c}{}_{ab}X_{c}\;.\;\;\;\; 
\]
It follows from (\ref{CC'}) that 
\begin{equation}
X_{a}(\alpha ^{b}{}_{\mu })=f^{b}{}_{ca}\;\alpha ^{c}{}_{\mu }\;.
\label{adjbehavior}
\end{equation}
This behavior characterizes $\alpha $ as a connection, or an adjoint-behaved 
$1-$form. It may seem that a derivative, vacuum term is missing, but we are
working on the bundle and the vacuum term only comes out when the connection
is pulled--back to spacetime by a section.

From (\ref{betaprime1}) and (\ref{K}) we obtain the expression for the
non--linearity indicator: 
\begin{equation}
\beta ^{a}{}_{\mu \nu }=\partial _{\mu }\alpha ^{a}{}_{\nu }-\partial _{\nu
}\alpha ^{a}{}_{\mu }+f^{a}{}_{bc}\alpha ^{b}{}_{\mu }\alpha ^{c}{}_{\nu }\;.
\label{betaprime2}
\end{equation}
Since in a direct product the extension is central, we must have 
\begin{equation}
\left[ X_{a},\beta ^{c}{}_{\mu \nu }X_{a}\right] =0,
\end{equation}
and consequently 
\begin{equation}
X_{a}(\beta ^{c}{}_{\mu \nu })=f^{c}{}_{ba}\beta ^{b}{}_{\mu \nu }\;.
\end{equation}
This condition, which can be equally obtained from (\ref{Jacobi2}),
says that also $\beta $ belongs to the adjoint representation of the
group whose generators are represented by $X_{a}$.

The above algebraic configuration is just the structure appearing in a gauge
theory, where $\beta$ is the field strength of the gauge potential $\alpha$.
The change of basis (\ref{change2}) corresponds to the covariant derivative
introduced in gauge theories by the minimal coupling prescription.

Gauge field dynamics can be obtained via the duality prescription: the
sourceless field equations are written just as the Bianchi identities,
but applied to the dual of the field strength.  This dual depends on
the metric.  Recall that, of Maxwell's equations, one pair is
metric--insensitive (they are Bianchi identities) while the other is
metric--dependent (they are the real dynamical equations).  In
principle, any metric which is preserved by the derivation will do,
but different metrics lead to inequivalent equations.  We simply
assume the existence of such a metric.  We obtain the field equations
for $\alpha$ by first finding the Jacobi identity for three fields
$X_{\mu},X_{\nu},X_{\rho}$ in algebra (\ref{dirprod}) -- which gives a
Bianchi identity -- and then applying the duality prescription.  The
Yang--Mills equations come out:
\begin{equation}
X_{\mu }\beta ^{a\mu \nu }=0\;.  \label{gauge}
\end{equation}

From the point of view of the theory of algebra extensions, the next
natural step would be to break the direct product in (\ref{dirprod})
by another change of basis,
\begin{equation}
X^{\prime }{}_{\mu }=X_{\mu }-\gamma ^{a}{}_{\mu }X_{a}
\label{Change3}
\end{equation}
and investigate the kind of physical theory the resulting
configuration can be associated to.  Expression (\ref{Change3}) leads
to the following commutation relations
\begin{eqnarray}
\left[ X^{\prime }{}_{\mu },X^{\prime }{}_{\nu }\right] &=&-\;\beta ^{\prime
a}{}_{\mu \nu }X_{a};  \nonumber \\
\left[ X^{\prime }{}_{\mu },X_{a}\right] &=&C^{\prime c}{}_{\mu
a}X_{c};\;\;\;\;\   \label{C''} \\
\left[ X_{a},X_{b}\right] &=&f^{c}{}_{ab}X_{c},\ \;\;\;\;\;\;\;  \nonumber
\end{eqnarray}
which just corresponds to the extended theory of the previous section. 
Now, it follows from (\ref{CC'}) that
\begin{equation}
X_{b}(\gamma ^{a}{}_{\mu })=f^{a}{}_{cb}\;\gamma ^{c}{}_{\mu
}+C^{\prime a}{}_{\mu b}.  \label{misbehavior1}
\end{equation}
Comparison with (\ref{adjbehavior}) shows that $C^{\prime a}{}_{\mu
b}$ measures the deviation from covariant behavior of the object
$\gamma ^{a}{}_{\mu }$ appearing in (\ref{Change3}).  With the help of
(\ref{change2}), we can express (\ref{Change3}) as
\begin{equation}
X^{\prime }{}_{\mu }=\partial _{\mu }-\sigma ^{a}{}_{\mu }X_{a}
\label{dervgeral1}
\end{equation}
with $\sigma ^{a}{}_{\mu }\equiv (\alpha ^{a}{}_{\mu }+\gamma
^{a}{}_{\mu })$.  We shall call (\ref{dervgeral1}) a {\em generalized
derivative}.  In fact we shall, from now on, give that name to each
derivative which is not the standard gauge--covariant derivative.  The
behavior of $\sigma $ under the group action is
\begin{equation}
X_{b}(\sigma ^{a}{}_{\mu })=f^{a}{}_{cb}\;\sigma ^{c}{}_{\mu }+C^{\prime
a}{}_{\mu b}\;.  \label{misbehavior2}
\end{equation}
The new non-linearity indicator $\beta ^{\prime a}{}_{\mu \nu }$ can be
obtained from (\ref{betaprime1}) using (\ref{K}) and (\ref{betaprime2}): 
\begin{equation}
\beta ^{\prime a}{}_{\mu \nu }=\partial _{\mu }\sigma ^{a}{}_{\nu }-\partial
_{\nu }\sigma ^{a}{}_{\mu }+f^{a}{}_{bc}\sigma ^{b}{}_{\mu }\sigma
^{c}{}_{\nu }-C^{\prime a}{}_{c\mu }\ \sigma ^{c}{}_{\nu }+\;C^{\prime
a}{}_{c\nu }\ \sigma ^{c}{}_{\mu }  \label{beta''}
\end{equation}
($C^{\prime a}{}_{b\mu }$ = - $C^{\prime a}{}_{\mu b}$). This is the general
expression for the deviation from linearity in the presence of an object
with behavior given by (\ref{misbehavior2}). The behavior of $\beta ^{\prime
}$ under the group action is fixed by the Jacobi identity (\ref{Jacobi2}),
replacing $Y_{\mu }$ by $X^{\prime }{}_{\mu }$ and $C$ by $C^{\prime }$.

Dynamics associated to algebra (\ref{C''}) is obtained by applying the
duality prescription in a way analogous to that leading to the Yang--Mills
equations. The Jacobi identity involving three fields $X^{\prime}{}_{\mu }$
in (\ref{C''}) is
$$%
X^{\prime}{}_{\mu}(\beta^{\prime a}{}_{\nu \sigma}) - C^{\prime 
a}{}_{c \mu} \beta^{\prime c}{}_{\nu \sigma} + 
X^{\prime}{}_{\sigma}(\beta^{\prime a}{}_{\mu \nu}) - C^{\prime 
a}{}_{c \sigma} \beta^{\prime c}{}_{\mu \nu} \; \; \; \; \; \; \; \; 
\; \; \; \; \; \; $$%
\begin{equation}%
  \; \; \; \; \; \; \; \; \; \; \; \; \; \; \; \; \; \; \; \; \; \; \; 
  \; \; \; \; \; \; \; \; \; \; \; \; \; \; \; \; \; \; \; \; \; \; 
  \;\; \; \; \; +X^{\prime}{}_{\nu}(\beta^{\prime a}{}_{\sigma \mu }) 
  - C^{\prime a}{}_{c \nu} \beta^{\prime c}{}_{\sigma \mu} = 0 \; .  
  \label{Jacobistill} \end{equation}%
Applying this expression to the dual of $\beta^{\prime a}{}_{\mu \nu}$, the
field equations turn out to be 
\begin{equation}
X^{\prime }{}_{\mu}\beta ^{\prime a\mu \nu }-C^{\prime a}{}_{d\mu}\beta
^{\prime d\mu \nu }=0\;.  \label{eq''}
\end{equation}
These equation are, of course, linked to the choice of $C^{\prime }$, which
is constrained by the Jacobi identity (\ref{Jacobi3}).

The set of commutators (\ref{C''}) can be obtained directly from (\ref
{config1}) by the basis change (\ref{dervgeral1}). The above two-step
procedure is, however, appropriate to show how it can be attained from the
algebraic scheme of a gauge theory. The 1-form $\sigma ^{a}{}_{\mu }$
appearing in the generalized derivative can be seen as a connection deformed
by the addition of a non--covariant form (Aldrovandi 2, 1991).

We can infer using (\ref{beta''}) in (\ref{eq''}) that a mass term for $%
\sigma $ can appear.  Thus, this second change of basis (or a change
of basis in a gauge configuration) leads to a theory with massive
vector fields which do not behave like connections.  This is what 
happens in the Weinberg-Salam model.

Another change of basis may be introduced as follows. Going back to (\ref
{Jacobistill}) we see that it has the form of a Bianchi identity for a still
more general, enlarged derivative $X^{\prime \ast }{}_{\mu }$, which can be
defined by its action on a indexed object $Z^{c}$ as 
\begin{equation}
X^{\prime \ast }{}_{\mu }(Z^{c})=X^{\prime }{}_{\mu }(Z^{c})-C^{\prime
c}{}_{a\mu }(Z^{a})\;.  \label{estrela}
\end{equation}

To be acceptable as a derivative, $X^{\prime *}{}_{\mu }$ must obey
the Leibniz rule, which leads to some interesting consequences.  For
example, for a scalar of type $Z^{a}Z_{a}$,
\begin{equation}
X^{\prime *}{}_{\mu }(Z^{a}Z_{a})=X^{\prime }{}_{\mu }(Z^{a}Z_{a}) \; .
\label{P1}
\end{equation}
For a lower--indexed object, 
\begin{equation}
X^{\prime *}{}_{\mu }(Z_{c})=X^{\prime }{}_{\mu }(Z_{c})+C^{\prime
e}{}_{c\mu }Z_{e} \; ,  \label{P2}
\end{equation}
and for a mixed product, 
\begin{equation}
X^{\prime *}{}_{\mu }(Z^{d}J_{c})=X^{\prime }{}_{\mu
}(Z^{d}J_{c})+C^{\prime e}{}_{c\mu }(Z^{d}J_{e})-C^{\prime d}{}_{e\mu
}(Z^{e}J_{c}) \; .  \label{P3}
\end{equation}

Expression (\ref{estrela}) leads to the commutators
\begin{equation}
\left[ X^{\prime \ast }{}_{\mu },X^{\prime \ast }{}_{\nu }\right]
(Z^{c})=-\beta ^{\prime a}{}_{\mu \nu }X_{a}(Z^{c})-R^{\prime
c}{}_{a\mu \nu }Z^{a}\;; \label{quiproquo1}
\end{equation}
\begin{equation}
\left[ X^{\prime \ast }{}_{\mu },X_{a}\right] (Z^{c})=X_{a}(C^{\prime
c}{}_{d\mu })Z^{d}\;, \label{possib}
\end{equation}
where $\beta ^{\prime a}{}_{\mu \nu }$ is given by (\ref{beta''}) and
\begin{equation}
R^{\prime c}{}_{a\mu \nu }=X^{\prime }{}_{\mu }C^{\prime c}{}_{a\nu
}-X^{\prime }{}_{\nu }C^{\prime c}{}_{a\mu }-C^{\prime c}{}_{b\mu
}C^{\prime b}{}_{a\nu }+C^{\prime c}{}_{b\nu }C^{\prime b}{}_{a\mu
}\;.
\label{curvtobe}
\end{equation}
The relation between $C^{\prime }$ and its algebraic derivative is given by
Jacobi identity (\ref{Jacobi3}).

Besides the same non-linear coefficient $\beta ^{\prime a}{}_{\mu \nu
}$ appearing in (\ref{C''}), the extra non--linear term $R^{\prime
c}{}_{a\mu \nu }$ turns up.  The relationship between these
coefficients is provided by the Jacobi identity for the fields
$X_{a}$,$X^{\prime }{}_{\mu }$,$X^{\prime }{}_{\nu }$:
\begin{equation}
X_{b}(\beta ^{\prime a}{}_{\mu \nu })+f^{a}{}_{bc}\beta ^{\prime c}{}_{\mu
\nu }+R^{\prime a}{}_{b\mu \nu }=0\;.  \label{betaerre}
\end{equation}

The dynamics corresponding to configuration (\ref{quiproquo1}) and (\ref
{possib}) is examined in the next two sections. 

%%%%%%%%%%%%%%%%%%%%%%%%%%%%%%%%%%

\section{Enlarging the geometry }
\label{recovering}
%%%%%%%%%%%%%%%%%%%%%%%%%%%%%%%%%%

In the fiber bundle structure, a local basis always exists (Cho, 1975) 
in which the commutation table takes up the form (\ref{dirprod}).  
This means that real geometry, or real bundles, only admit quantities 
behaving as connections.  Extended field algebras involve an object 
behaving differently.  We endeavor now to move a little beyond the 
strictly geometric canvas, by finding which properties can still be 
retained in the presence of such a misbehaving element and which 
requirements should be imposed if we insist in remaining as near as 
possible to usual geometry.

Firstly, we would like to relate the new objects to gravitation, and $%
R^{\prime c}{}_{a\mu \nu }$ would bear some resemblance to a curvature
written in the basis $\{X^{\prime }{}_{\nu }\}$ {\em if} $C^{\prime }$ were
a connection. However, (\ref{curvtobe}) is not the correct expression for a
curvature. A term involving a contraction of the basis anholonomy
coefficient with the connection is missing. Furthermore, the $\beta ^{\prime
a}{}_{\mu \nu }$ term in (\ref{quiproquo1}) should be a torsion, or an
anholonomy, but is not: for that, it should have values along $X^{\prime
\ast }{}_{\mu }$.

 Under the assumption that the dimensions of the two algebras are the
 same, a solution to these problems comes by the following
 considerations.  The vector spaces underlying two algebras of the
 same finite dimension are isomorphic (we insist: only as vector
 spaces).  The isomorphism can be realized by a mapping $H$ between
 them, such as
\begin{equation}
X^{\prime }{}_{\mu }=H^{a}{}_{\mu }X_{a}.  \label{aga}
\end{equation}
The mapping described by $H^{a}{}_{\mu }$ should be invertible.  If we
have spacetime in mind, the group should be itself 4-dimensional.  The
isomorphism in view would actually be between the tangent spaces, and
should hold at each point of the manifold.  Provided some reasonable
differentiability conditions are met, the set $\{H^{a}{}_{\mu }\}$
will be similar to a tetrad field.  We shall use for the inverse the
usual tetrad notation, so that $ H^{a}{}_{\mu }H_{b}{}^{\mu }=\delta
_{b}^{a}$ and $H^{a}{}_{\mu }H_{a}{}^{\nu }=\delta _{\mu }^{\nu }$. 
Applying (\ref{aga}) to the second commutator in (\ref{C''}) we obtain
\begin{equation}
X_{a}H^{d}{}_{\mu }=f^{d}{}_{ca}H^{c}{}_{\mu }-C^{\prime d}{}_{\mu
a}\;.
\label{h2}
\end{equation}
A brief calculation leads to 
\begin{equation}
\left[ X^{\prime }{}_{\mu },X^{\prime }{}_{\nu }\right] =-\beta ^{\prime
\rho }{}_{\mu \nu }X^{\prime }{}_{\rho }\;,
\end{equation}
with 
\begin{equation}
\beta ^{\prime \rho }{}_{\mu \nu }=\beta ^{\prime a}{}_{\mu \nu
}H_{a}{}^{\rho }\;,
\end{equation}
showing ($-\beta^{\prime \rho}{}_{\mu \nu}$) in the role of the
non--holonomy coefficient for the basis $\{X^{\prime }{}_{\mu }\}$.

Taking (\ref{aga}) into (\ref{estrela}) we obtain the relation between $%
X_{a} $ and $X^{\prime *}{}_{\mu }$ : 
\begin{equation}
X_{a}Z^{c}=H^{\mu }{}_{a}(X^{\prime *}{}_{\mu }Z^{c}+C^{\prime c}{}_{b\mu
}Z^{b})\;.  \label{chiza}
\end{equation}
The commutator (\ref{quiproquo1}) can then be rewritten as 
\begin{equation}
\left[ X^{\prime *}{}_{\mu },X^{\prime *}{}_{\nu }\right] (Z^{c})=-\beta
^{\prime \rho }{}_{\mu \nu }X^{\prime *}{}_{\rho }Z^{c}-{\mathcal R}
{}^{c}{}_{a\mu \nu }Z^{a}\;,  \label{quiproquo3}
\end{equation}
where now 
\begin{equation}
{}{\mathcal R}^{c}{}_{a\mu \nu }=X^{\prime }{}_{\mu }C^{\prime c}{}_{a\nu
}-X^{\prime }{}_{\nu }C^{\prime c}{}_{a\mu }-C^{\prime c}{}_{b\mu }C^{\prime
b}{}_{a\nu }+C^{\prime c}{}_{b\nu }C^{\prime b}{}_{a\mu }+\beta ^{\prime
\rho }{}_{\mu \nu }C^{\prime c}{}_{a\rho }\;.  \label{curvtobe2}
\end{equation}
This is the correct expression of the curvature of a connection $C^{\prime }$
in basis $\{X^{\prime }{}_{\mu }\}$ (Nakahara, 1990). It can be shown that the
commutator in (\ref{quiproquo3}), if applied to a mixed object with internal
and external indices, only acts on those internal.

Let us examine some more properties of the candidate--connection
$C^{\prime}$.  Taking $X_{a}=H^{\mu }{}_{a}X^{\prime }{}_{\mu }$ into
the second commutator of (\ref{C''}), we obtain
\begin{equation}
C^{\prime b}{}_{a\mu }=H^{b}{}_{\lambda }C^{\prime \lambda }{}_{\nu
\mu }H_{a}{}^{\nu }+H_{a}{}^{\nu }X^{\prime }{}_{\mu }(H^{b}{}_{\nu
}),
\label{hah}
\end{equation}
which shows that $C^{\prime }$ behaves, under the the action of
$H^{\mu }{}_{a}$, as a connection of the linear group would behave
under a tetrad, with $C^{\prime \lambda }{}_{\nu \mu }=\beta ^{\prime
\lambda }{}_{\mu \nu }$.  A curious consequence is that
\begin{equation}
X^{\prime *}{}_{\lambda} \beta ^{\prime \rho}{}_{\mu \nu } =
X^{\prime}{}_{\lambda} \beta ^{\prime \rho}{}_{\mu \nu } \; .
\label{property}
\end{equation}
The torsion tensor is $T^{\rho }{}_{\mu \nu }=-\beta ^{\prime \rho
}{}_{\mu \nu }$.  From (\ref{h2}) and (\ref{hah}) it can be written as
\begin{equation}
T^{\rho }{}_{\mu \nu }=H_{a}{}^{\rho }(X^{\prime }{}_{\mu
}H^{a}{}_{\nu }-X^{\prime }{}_{\nu }H^{a}{}_{\mu
}+f^{a}{}_{bc}H^{b}{}_{\mu }H^
{c}{}_{\nu
}).  \label{newtorsion}
\end{equation}
The deformed Yang-Mills field strength acquires the rank of a torsion. Due
to the last term in (\ref{newtorsion}), we would better call $T^{\rho
}{}_{\mu \nu }$ a ``generalized torsion tensor''. It reduces to usual
torsion in the abelian case (Aldrovandi 2, 1991). One should remember that usual
torsion is a 2-form with values in the algebra of translation generators. In
the present case, torsion has values in the assumed gauge group algebra,
which is non abelian. This is the origin of the extra term.

As with usual tetrads, the $H^{a}{}_{\mu }$'s can be used to transmute
indices from the gauge algebra to spacetime and vice--versa. However, due to
the presence of non-covariant objects, the usual properties do not follow
automatically --- every one must be verified by direct calculation. For
example, computation gives, for the curvature, just what we would expect
from a tensorial object, 
\begin{equation}
{\mathcal R}^{\rho }{}_{\sigma \mu \nu }=X^{\prime }{}_{\mu }C^{\prime \rho
}{}_{\sigma \nu }-X^{\prime }{}_{\nu }C^{\prime \rho }{}_{\sigma \mu
}-C^{\prime \rho }{}_{\alpha \mu }C^{\prime \alpha }{}_{\sigma \nu
}+C^{\prime \rho }{}_{\alpha \nu }C^{\prime \alpha }{}_{\sigma \mu }+\beta
^{\prime \gamma }{}_{\mu \nu }C^{\prime \rho }{}_{\sigma \gamma }\;.
\label{sptimeCurvature}
\end{equation}
As in (\ref{property}), it happens that 
\begin{equation}
X^{\prime \ast }{}_{\lambda }{\mathcal R}^{\rho }{}_{\sigma \mu \nu }=X^{\prime
}{}_{\lambda }{\mathcal R}^{\rho }{}_{\sigma \mu \nu }\;.  \label{property2}
\end{equation}
It is important to notice that the enlarged derivative, once acting on
objects with the indices transmuted to spacetime indices, changes its form.
As happens with covariant derivatives, it will take a different aspect when
acting on objects with one or two indices. The simplest way to discover its
form is to read it from the Bianchi identities, as we shall do below.

%%%%%%%%%%%%%%%%%%%%%%%%%%%%%%%%

\section{Approaching a gravitational model}
\label{EqsandIdents}
%
%%%%%%%%%%%%%%%%%%%%%%%%%%%%%%%%

We have in the previous section succeeded in obtaining (i) an
anholonomy or torsion term in the commutator and (ii) the correct
expression for the curvature in a non-holonomic basis.  We shall in
what follows show that two other geometrical landmarks also hold: the
two Bianchi identities for linear connections.  Despite their purely
geometrical character, Bianchi identities are, both in gauge theories
and in General Relativity, intimately related to dynamics, so that we
shall also comment on the field equations.  The procedure adopted here
parallels those theories.  The field equations are obtained by
applying the duality prescription to the sole Bianchi identity present
in the gauge sector.  In the gravity sector, a contracted Bianchi
identity is used to recognize which expression is to be identified to
the source current.  Properties (\ref{property}) and (\ref{property2})
hold in general when we derive objects with external indices only. 
Thus, the metric $g_{\alpha \beta}$ used in (\ref{eq''}) can be used
in the following.  We see from (\ref{aga}) that, if preserved by
$X^{\prime *}{}_{\mu }$, it will be also gauge invariant.  Recognizing
(\ref{estrela}) in the field equation (\ref {eq''}) and adding a
source current, we arrive at
\begin{equation}
X^{\prime *}{}_{\mu } \beta ^{\prime a \mu \nu}=J^{a\nu } \; . 
\label{m1}
\end{equation}
As the deformed Yang--Mills field coincides with torsion, this
equation fixes the dynamics for both.  Applying $X^{\prime *}{}_{\nu
}$ to this equation, a rather surprising result turns up:
\begin{equation}
X^{\prime *}{}_{\nu }J^{a\nu }=0 \; .
\end{equation}
This ``current conservation'' shows that some invariance must still be
at work, although its meaning is not clear.  Notice that the
commutation relations, the new covariant derivatives and the dynamics
of $\sigma ^{c}{}_{\mu }$ have all been constructed or obtained in the
respect of the Jacobi identities which are, for tangent vector fields,
integrability conditions.  Once also the duality symmetry is supposed
to hold, some invariance is to be expected.

Which kind of gravitational model would turn up?  Using
(\ref{curvtobe2}), equation (\ref{betaerre}) can be written as
\begin{equation}
X_{a}\beta^{\prime c}{}_{\mu \nu }+f^{c}{}_{a e}\beta^{\prime e}{}_{\mu
\nu }=-{\mathcal R}^{c}{}_{a \mu \nu }+H^{\rho}{}_{d}\beta^{\prime
d}{}_{\mu \nu }C^{\prime c}{}_{a\rho } \; , \label{g4}
\end{equation}
which presents ${\mathcal R}^{b}{}_{a \mu \nu }$ as an effect of $\beta$'s
non-covariance.  Applying $H^{\alpha }{}_{c}H^{a}{}_{\sigma}$, we find
\begin{equation}
{\mathcal R}^{\alpha }{}_{\sigma \mu \nu }+X^{\prime *}{}_{\sigma }(\beta
^{\prime \alpha }{}_{\mu \nu })=0 \;.  \label{g8}
\end{equation}
The Ricci tensor is not symmetric, 
\begin{equation}
{\mathcal R}_{\sigma \nu}+X^{\prime *}{}_{\sigma }(\beta ^{\prime \alpha
}{}_{\alpha \nu })=0 \; ,  \label{ricci}
\end{equation}
which is to be expected in the presence of torsion. The gravitational sector
would be close to an Einstein-Cartan model, but with dynamical torsion.
Combined with (\ref{g8}), (\ref{sptimeCurvature}) leads to 
$$%
X^{\prime }_{\lambda} (\beta^{\prime \rho }{}_{\nu \mu}) + X^{\prime 
}_{\mu} (\beta^{\prime \rho }{}_{\lambda \nu}) + X^{\prime }_{\nu} 
(\beta^{\prime \rho }{}_{\mu} \lambda) = \; \; \; \; \; \; \; \; \; \; 
\; \; \; \; \; \; \; \; \; \; \; \; \; \; \; \; \; \; \; \; \; \; \; %
$$%
\begin{equation}%
\; \; \; \; \; \; \; \; \; \; \; \; \; \; \; \; \; \; \; \; \; \; \; 
\; \; \; \; \; \; \; \; \; \; \; \; \; \; \; \; \; \; \; \; \; - \ 
\beta ^{\prime \alpha}{}_{\mu \nu} \beta ^{\prime \rho }{}_{\lambda 
\alpha} - \beta ^{\prime \alpha}{}_{\lambda \mu} \beta^{\prime \rho 
}{}_{\nu \alpha} - \beta ^{\prime \alpha}{}_{\nu \lambda} \beta 
^{\prime \rho }{}_{ \mu \alpha} \; .\label{yetone} \end{equation}%

Let us now show that the two Bianchi identities have the same formal
aspect they have in usual geometry.  For that, we start by calculating
of the Jacobi identity for $X^{\prime *}{}_{\mu },X^{\prime *}{}_{\nu
}$ and $X^{\prime *}{}_{\lambda}$,
$$%
\left[ X^{\prime *}{}_{\lambda },\left[ X^{\prime *}{}_{\mu 
},X^{\prime *}{}_{\nu }\right] \right] (Z^{c})+ \left[ X^{\prime 
*}{}_{ \nu},\left[ X^{\prime *}{}_{\lambda },X^{\prime *}{}_{\mu 
}\right] \right] (Z^{c}) \; \; \; \; \; \; \; \; \; \; \; \; \; \; \; 
\; \; \; \; \; \; \; %
$$%
\begin{equation}%
\; \; \; \; \; \; \; \; \; \; \; \; \; \; \; \; \; \; \; \; \; \; \; 
\; \; \; \; \; \; \; \; \; \; \; \; \; \; \; \; \; \; \; \; \; \; \; 
\; \; \; \; \; \; \; \; \; + \left[ X^{\prime *}{}_{\mu },\left[ 
X^{\prime *}{}_{ \nu},X^{\prime *}{}_{\lambda }\right] \right] (Z^{c}) 
= 0 \; .  \label{Jacobiyet} \end{equation}%
We first obtain one of the three cyclic terms: 
$$%
\left[ X^{\prime *}{}_{\lambda },\left[ X^{\prime *}{}_{\mu 
},X^{\prime *}{}_{\nu }\right] \right] (Z^{c}) = \;\left[ \beta 
^{\prime \rho }{}_{\mu \nu } \beta ^{\prime \alpha }{}_{\lambda \rho} 
- X^{\prime *}{}_{\lambda } \beta ^{\prime \alpha }{}_{\mu \nu} 
\right] X^{\prime *}{}_{\alpha } (Z^{c})\; \; \; \; \; 
$$%
$$%
\; \; \; \; \; \; \; \; \; \; \; \; \; \; \; \; \; \; \; \; \; \; \; 
\; \; \; \; \; \; \; \; \; \; \; \; \; \; \; \; \; \; \; \; \; \; \; 
\; \; \; \; \; \; \; \; \; \; \; \; \; \; \; \; \; \; \; \; \; \; - \left[ 
X^{\prime *}{}_{\lambda } {\mathcal R}^{c}{}_{a \mu \nu} - \beta 
^{\prime \rho }{}_{\mu \nu } {\mathcal R}^{c}{}_{a \rho \lambda} 
\right] Z^{a} .
$$%

We can here read the enlarged derivative acting on an object with one
transmuted index.  The expression inside the first bracket in the
right-hand-side is, up to the sign, equal to $X^{\prime}{}_{\lambda }
\beta ^{\prime \alpha }{}_{\mu \nu} - C^{\prime \alpha }{}_{\rho
\lambda} \beta ^{\prime \rho }{}_{\mu \nu}$.  This is the enlarged
derivative acting on $\beta ^{\prime \alpha }{}_{\mu \nu}$.

Applying (\ref{g8}), the tensorial character of ${\mathcal R}$ and
$X^{\prime \ast }{}_{\lambda }(H_{a}{}^{\sigma })=H_{a}{}^{\delta
}C^{\prime \sigma }{}_{\delta \lambda }$, this expression can be
rewritten as
\[
\left[ X^{\prime \ast }{}_{\lambda },\left[ X^{\prime \ast }{}_{\mu
},X^{\prime \ast }{}_{\nu }\right] \right] (Z^{c})=\left[ \beta
^{\prime \rho }{}_{\mu \nu }\beta ^{\prime \alpha }{}_{\lambda \rho
}+{\mathcal R} ^{\alpha }{}_{\lambda \mu \nu }\right] X^{\prime \ast
}{}_{\alpha }(Z^{c})
\]
\[
+H^{c}{}_{\alpha }H_{a}{}^{\sigma }\left[ {\mathcal R}^{\rho }{}_{\sigma
\mu \nu }\beta ^{\prime \alpha }{}_{\lambda \rho }-{\mathcal R}^{\alpha
}{}_{\rho \mu \nu }\beta ^{\prime \rho }{}_{\lambda \sigma }-{\mathcal
R}^{\alpha }{}_{\sigma \rho \lambda }\beta ^{\prime \rho }{}_{\mu \nu
}-X^{\prime \ast }{}_{\lambda }{\mathcal R}^{\alpha }{}_{\sigma \mu \nu
}\right] Z^{a}.
\]
Identity (\ref{Jacobiyet}) becomes then 
\begin{eqnarray}
X^{\prime \ast }{}_{\alpha }(Z^{c})\left[ \beta ^{\prime \rho }{}_{\mu
\nu }\beta ^{\prime \alpha }{}_{\lambda \rho }+\beta ^{\prime \rho
}{}_{\lambda \mu }\beta ^{\prime \alpha }{}_{\nu \rho }+\beta ^{\prime
\rho }{}_{\nu \lambda }\beta ^{\prime \alpha }{}_{\mu \rho }+{\mathcal
R}^{\alpha }{}_{\lambda \mu \nu }+{\mathcal R}^{\alpha }{}_{\nu \lambda
\mu }+{\mathcal R}^{\alpha }{}_{\mu \nu \lambda }\right] \nonumber \\
=-H^{c}{}_{\alpha }H_{a}{}^{\sigma }Z^{a}{\bf \{}X^{\prime \ast
}{}_{\nu } {\mathcal R}^{\alpha }{}_{\sigma \lambda \mu }-C^{\prime \alpha
}{}_{\rho \nu } {\mathcal R}^{\rho }{}_{\sigma \lambda \mu }+C^{\prime
\rho }{}_{\sigma \nu } {\mathcal R}^{\alpha }{}_{\rho \lambda \mu }-\beta
^{\prime \rho }{}_{\lambda \nu }{\mathcal R}^{\alpha }{}_{\sigma \rho \mu
}\;\;\;\; \nonumber \\
+X^{\prime \ast }{}_{\lambda }{\mathcal R}^{\alpha }{}_{\sigma \mu \nu
}-C^{\prime \alpha }{}_{\rho \lambda }{\mathcal R}^{\rho }{}_{\sigma \mu
\nu }+C^{\prime \rho }{}_{\sigma \lambda }{\mathcal R}^{\alpha }{}_{\rho
\mu \nu }-\beta ^{\prime \rho }{}_{\mu \lambda }{\mathcal R}^{\alpha
}{}_{\sigma \rho \nu }\;\;\;\;\;\;\;\;\;\;\;\;\;\;\;\;\;\;\;\;
\nonumber \\
+X^{\prime \ast }{}_{\mu }{\mathcal R}^{\alpha }{}_{\sigma \nu \lambda
}-C^{\prime \alpha }{}_{\rho \mu }{\mathcal R}^{\rho }{}_{\sigma \nu
\lambda }+C^{\prime \rho }{}_{\sigma \mu }{\mathcal R}^{\alpha }{}_{\rho
\nu \lambda }-\beta ^{\prime \rho }{}_{\mu \nu }{\mathcal R}^{\alpha
}{}_{\sigma \rho \lambda }{\bf \}} =0\;.\;\;\;\;\;\; \label{monster}
\end{eqnarray}
With the term proportional to $X^{\prime \ast }{}_{\alpha }(Z^{c})$ in
view, we calculate
\[
{\mathcal R}^{\alpha }{}_{\lambda \mu \nu }+{\mathcal R}^{\alpha }{}_{\nu \lambda
\mu }+{\mathcal R}^{\alpha }{}_{\mu \nu \lambda }+\beta ^{\prime \rho }{}_{\mu
\nu }\beta ^{\prime \alpha }{}_{\lambda \rho }+\beta ^{\prime \rho
}{}_{\lambda \mu }\beta ^{\prime \alpha }{}_{\nu \rho }+\beta ^{\prime \rho
}{}_{\nu \lambda }\beta ^{\prime \alpha }{}_{\mu \rho }= 
\]
\[
2\left[ X_{\mu }^{\prime }C^{\prime \alpha }{}_{\lambda \nu }+X_{\nu
}^{\prime }C^{\prime \alpha }{}_{\mu \lambda }+X_{\lambda }^{\prime
}C^{\prime \alpha }{}_{\nu \mu }-\beta ^{\prime \rho }{}_{\mu \nu }\beta
^{\prime \alpha }{}_{\lambda \rho }-\beta ^{\prime \rho }{}_{\lambda \mu
}\beta ^{\prime \alpha }{}_{\nu \rho }-\beta ^{\prime \rho }{}_{\nu \lambda
}\beta ^{\prime \alpha }{}_{\mu \rho }\;\right] \;. 
\]
The term inside the brackets vanishes by the Jacobi identity (\ref
{Jacobistill}) with all the indices in spacetime. The left--hand side is the
factor of $X^{\prime \ast }{}_{\alpha }(Z^{c})$ in the first term of (\ref
{monster}), which consequently vanishes too. The remaining content of (\ref
{monster}) is the vanishing of the term proportional to $Z^{a}$, whose
meaning we examine in the following. Let us notice before that the above
left--hand side has another interest: the fact that it is zero, combined
with (\ref{yetone}), results in 
\begin{equation}
{\mathcal R}^{\alpha }{}_{\lambda \mu \nu }+{\mathcal R}^{\alpha }{}_{\nu \lambda
\mu }+{\mathcal R}^{\alpha }{}_{\mu \nu \lambda }=X_{\lambda }^{\prime }(\beta
^{\prime \rho }{}_{\nu \mu })+X_{\mu }^{\prime }(\beta ^{\prime \rho
}{}_{\lambda \nu })+X_{\nu }^{\prime }(\beta ^{\prime \rho }{}_{\mu \lambda
})\;.  \label{Bianchifirst}
\end{equation}
As $\beta ^{\prime \rho }{}_{\mu \nu }$ is the torsion, this is just the
expression of the first Bianchi identity to which linear connections 
submit (Kobayashi and Nomizu, 1963).

To analyse the vanishing of the term proportional to $Z^{a}$, it is
convenient to read in (\ref{monster}) itself the form of the enlarged
derivative in the non-holonomic basis $\{X_{\mu }^{\prime }\}$ when
applied to an object with two transmuted indices, like ${\mathcal
R}^{\alpha }{}_{\lambda \mu \nu }$.  It has the same expression of the
usual covariant derivative:
\begin{equation}
D_{\nu }{\mathcal R}^{\alpha }{}_{\sigma \lambda \mu }=X_{\nu }^{\prime
}{\mathcal R} ^{\alpha }{}_{\sigma \lambda \mu }-C^{\prime \alpha
}{}_{\rho \nu }{\mathcal R} ^{\rho }{}_{\sigma \lambda \mu }+C^{\prime
\rho }{}_{\sigma \nu }{\mathcal R} ^{\alpha }{}_{\rho \lambda \mu }-\beta
^{\prime \rho }{}_{\lambda \nu }{\mathcal R}^{\alpha }{}_{\sigma \rho \mu
}\;,
\end{equation}
where use has beeen made of (\ref{property2}). The identity then reduces to 
\begin{equation}
D_{\nu }{\mathcal R}^{\alpha }{}_{\sigma \lambda \mu }+D_{\lambda }{\mathcal R}%
^{\alpha }{}_{\sigma \mu \nu }+D_{\mu }{\mathcal R}^{\alpha }{}_{\sigma
\nu \lambda }=0\;, \label{Bianchisecond}
\end{equation}
which has the form of the second Bianchi identity. 

We now follow a path which parallels that used in General Relativity
to identify the geometrical object appearing in the field equation,
analogous to the Einstein tensor.  Contracting first $\alpha $ with
$\lambda $ and then using the preserved metric $g_{\alpha \beta }$ to
contract the remaining indices, we get
\[
D_{\nu }{\mathcal R}^{\mu }{}_{\mu }+D_{\alpha }{\mathcal R}^{\alpha \mu
}{}_{\mu \nu }+D_{\mu }{\mathcal R}^{\mu }{}_{\nu }=0\;.
\]
This contracted Bianchi identity takes the form 
\begin{equation}
D_{\alpha }G^{\alpha \sigma }=0
\end{equation}
provided we define 
\begin{equation}
G^{\alpha \sigma }={\mathcal R}^{\alpha \sigma }-g^{\alpha \sigma }{\mathcal
R} -g^{\sigma \nu }{\mathcal R}^{\alpha \mu }{}_{\mu \nu }\;.
\end{equation}
This expression would lead to an object quite similar to the Einstein
tensor if ${\mathcal R}^{\alpha \beta }{}_{\lambda \mu }$ were
antisymmetric in the first two indices.

We have above obtained the two Bianchi identities of usual geometry with
torsion. To recover all the features of a real geometry the only missing
point is the direct product of the vector field algebras. We see in (\ref
{possib}) the possibility of recovering the direct product by setting $
X_{a}(C^{\prime c}{}_{d\mu })$ $Z^{d}=0$, which includes the
invariance of $ C^{\prime }$ under the group action,
\begin{equation}
X_{a}(C^{\prime c}{}_{d\mu })=0\;.  \label{Cinvariant}
\end{equation}
This would mean a constant $C^{\prime a}{}_{b\mu }$, but not a
constant $ C^{\prime \rho }{}_{\mu \nu }$, so that the curvature would
keep its general form (\ref{sptimeCurvature}).  It is important to
notice that such a condition to establish the direct product could
only be realized because we have made the change of basis
(\ref{estrela}).  It could not be done inside the extended gauge
theory, since we wanted to preserve the misbehaving elements.

The validity of the scheme is restricted to gauge groups with the
dimension of spacetime.  In consequence the extended gauge scheme,
besides describing a theory with massive fields that do not behave as
connections, also describes a theory for a group with the dimension of
spacetime.  If we take this dimension equal to four, a group that
could be chosen is the $SU(2)\otimes U(1)$ of the Weinberg Salam
model.  The condition in the dimension of the group guarantees the
existence of the mapping $H_{a}{}^{\nu }$ and its inversibility.  The
introduction of $H$ allowed us to recover the usual geometric
interpretation of curvature and torsion.  Significantly enough, the
behavior of $C^{\prime }$ is the same as that of an external
connection.  The gravitational sector would exhibit curvature and see
the deformed gauge field as a torsion, with dynamics given
respectively by (\ref{ricci}) and (\ref{m1}).

%%%%%%%%%%%%%%%%%%%%%%%%%%%%%%%%

\section{Open questions and final remarks}

%%%%%%%%%%%%%%%%%%%%%%%%%%%%%%%%

We have seen how, when a gauge potential ceases to behave like a
connection, the bundle picture of gauge theories becomes shaky.  The
arena appropriate to discuss the new situation is no more bundle
theory, but the theory of Lie algebra extensions, which allows for the
modified local commutators coming to the fore.  The object measuring
the breaking of the bundle scheme is reminiscent of a linear,
external, spacetime connection.  A new, non-covariant, generalized
derivative emerges naturally which is analogous to that appearing in
electroweak theory in the presence of a gravitational field.  This
suggests a link between electroweak interactions and gravitation.  The
suggestion is strengthened by a dimensional coincidence: spacetime and
the Lie algebra of the electroweak group are both 4-dimensional and,
as vector spaces, isomorphic.  This isomorphism can be realized by a
tetrad--like field $H$ which, once introduced, reorganizes the whole
picture.  Objects corresponding to the curvature and the torsion of
the candidate linear connection turn up at the right places in the
commutation relations and obey formally the two Bianchi identities of
Differential Geometry.  The broken gauge field strength appears in the
role of torsion.  The dynamical equations obtained correspond formally
to a broken gauge model on a spacetime endowed with
curvature.\label{final}

We are far from having solved all the questions raised by the 
approach.  The crucial, obvious problem which remains unsolved is that 
of the necessary index transmutation.  We do obtain quantities {\em 
resembling} a connection, a curvature and a torsion by their behavior.  
The connections related to gravitation are, however, related to the 
Lorentz group.  This means that, instead of our internal indices, we 
should have indices related to some vector or tensor representation of 
the Lorentz group.  This is clear in the case of real tetrad fields, 
which are Lorentz vectors.  Our latin indices should be somehow 
transformed into Lorentz indices before we can really speak of 
gravitation.  This question is not easy to answer in a satisfactory 
way.  What we can do by now is to speculate on possible origins for 
such a transmutation.  A first point to look at is the definition of 
$H$.  We have assumed equal dimensions to avoid an ill-defined mapping 
and this has led to the transformation (\ref{hah}) of $C^{\prime}$, 
which takes an object with internal indices into an object exclusively 
``external''.  But the fact remains that the original group has 
nothing to do with spacetime.  When the gauge group is the group of 
spacetime translations $T^{3,1}$, $H$ reduces to the usual vierbeine 
fields, which appear quite naturally (Aldrovandi 2, 1991).  In that case 
$C^{\prime}$ turns up as a true connection for the linear or Lorentz 
group, with a Riemann curvature and an additional torsion.  However, 
translation generators are Lorentz vectors and, in a sense, 
``external'' from the start.  We mention in passing that the Lorentz 
group does not affect spacetime directly, but through a 
representation, the vector representation in the case.  It could 
happen that the group supposed above -- say, the group of the 
electroweak interactions -- come to do the same, so that the 
relationship to spacetime come to be realized through an intermediate, 
``interface'' representation.  This will depend on the available 
representations of the gauge group.  The group of electroweak 
interactions is at present under study.

A point worth mentioning concerns universality.  It is true that 
gravitation is the only universal interaction.  However, the 
electroweak interaction presents a large amount of universality.  
Though with different strengths, all particles (except possibly the 
gluons) couple to it.

For the time being, the only
positive clue we have to the possibility of transmutation is the
appearence of torsion in expressions like (\ref{quiproquo3}). 
Torsion is specifically external, an effect of soldering which is
absent in purely internal gauge bundles.  Even when it vanishes, it is
responsible for the presence of two --- instead of only one ---
Bianchi identities.  Another point worth remembering is that our
approach is, up to now, purely classical.  It is possible that
transmutation come as a quantum effect.  Indeed, getting ``spin from isospin'' has been studied in the seventies (Jackiw and Rebbi, 1976; 
Hasenfratz and `t Hooft, 1976; Goldhaber, 1976) as an
instanton--induced transmutation of exactly the required kind.
What we have done here has been to leave this question aside and
investigate the purely formal aspects of the approach, to see whether
it presents points enticing enough to justify further study.  We think
the results are highly positive.

\section*{Acknowledgments}

The authors are most grateful to FAPESP (S\~ao Paulo, Brazil), for
financial support.

\section*{References}

\noindent Aldrovandi 1 R., (1991). J. Math. Phys. \textbf{32} 2503.
 
\noindent Aldrovandi 2 R., (1991). Phys. Lett. \textbf{A155} 459.

\noindent Aldrovandi R., (1995).  \textit{Gauge Theories, and Beyond} 
in Barret, T.W., and Grimes, D.M.(eds.), \textit{Advanced 
Electromagnetism: Foundations, Theory and Applications}, World 
Scientific, Singapore, pp.  3-51.

\noindent Aldrovandi, R., and Pereira, J.G., (1995). \textit{An Introduction to Geometric Physics}, World Scientific, Singapore.

\noindent Cho, Y.M., (1975). J. Math. Phys. \textbf{16} 2029.

\noindent Daniel, M., and Viallet, C.M., (1980). Rev. Mod. Phys. \textbf{52} 175.

\noindent Gaillard, M.K., Grannis, P.D., and Sciulli, F.J., (1999).  
Rev. Mod. Phys. \textbf{71} S96.

\noindent Goldhaber, A.S., (1976). Phys.Rev.Lett. \textbf{36} 1122. 

\noindent Hasenfratz, P., and 't Hooft, G., (1976). Phys. Rev. Lett. \textbf{36} 1119.

\noindent Jackiw, R., and Rebbi, C., (1976). Phys. Rev. Lett. \textbf{36} 1116. 

\noindent Kobayashi, S., and Nomizu, K., (l963).  \textit{Foundations 
of Differential Geometry}, Interscience, New York, 1st vol.

\noindent Nakahara, M., (1990). \textit{Geometry, Topology and Physics}, IOP, Bristol.

\noindent Trautman, A., (1970). Rep. Math. Phys. \textbf{1} 29.

\noindent Wu, T.T., and Yang, C.N., (1975). Phys. Rev. \textbf{D12} 3845.

\end{document}